\documentclass[twocolumn,showpacs,preprintnumbers,amsmath,amssymb]{revtex4}

\usepackage{graphicx}
\usepackage{rotating}

\begin{document}

\draft

\title{Collective excitations of a degenerate gas at the BEC-BCS crossover}

\author{M. Bartenstein,$^{1}$ A. Altmeyer,$^{1}$ S. Riedl,$^{1}$ S. Jochim,$^{1}$ C. Chin,$^{1}$
J. Hecker Denschlag,$^{1}$ and R. Grimm$^{1,2}$}

\address{$^{1}$Institut f\"ur Experimentalphysik, Universit\"at Innsbruck, Technikerstra{\ss}e 25, 6020 Innsbruck,
Austria\\$^{2}$Institut f\"ur Quantenoptik und Quanteninformation,
\"Osterreichische Akademie der Wissenschaften, 6020 Innsbruck,
Austria}

\date{\today}

\pacs{34.50.-s, 05.30.Fk, 39.25.+k, 32.80.Pj}

\begin{abstract}
We study collective excitation modes of a fermionic gas of $^6$Li
atoms in the BEC-BCS crossover regime. While measurements of the
axial compression mode in the cigar-shaped trap close to a
Feshbach resonance confirm theoretical expectations, the radial
compression mode shows surprising features. In the strongly
interacting molecular BEC regime we observe a negative frequency
shift with increasing coupling strength.
In the regime of a strongly interacting Fermi gas, an abrupt
change in the collective excitation frequency occurs, which may be
a signature for a transition from a superfluid to a collisionless
phase.
\end{abstract}

\maketitle

\narrowtext

The crossover from a Bose-Einstein condensate (BEC) to a
Bardeen-Cooper-Schrieffer (BCS) superfluid has for decades
attracted considerable attention in many-body theory
\cite{becbcs}. Bose-Einstein condensates of molecules formed by
fermionic atoms of $^6$Li and $^{40}$K \cite{li2becinn, k2bec,
li2becmit, li2paris} provide a unique system to experimentally
explore this BEC-BCS crossover. In such ultracold gases
magnetically tuned scattering resonances, known as Feshbach
resonances, allow to control and vary the interaction strength
over a very broad range. Recent experiments have entered the
crossover regime and yield results on the interaction strength by
measuring the cloud size \cite{markus} and expansion energy
\cite{li2paris}. Moreover, two experiments
\cite{jin04,ketterlenew} have demonstrated the condensed nature of
fermionic atom pairs in the crossover regime.

Important questions are related to superfluidity in the crossover
regime \cite{holland}. When a molecular BEC is converted into an
ultracold Fermi gas \cite{markus}, one can expect ultralow
temperatures and superfluidity to extend far into the Fermi gas
regime \cite{carr}. Detection tools to probe superfluidity in this
regime are therefore requested. The investigation of collective
excitation modes \cite{stringari96} is well established as a
powerful method to gain insight into the physical behavior of
ultracold quantum gases in different regimes of Bose
\cite{stringaribook} and Fermi gases \cite{excite}. A recent paper
\cite{stringari04} points out an interesting dependence of the
collective frequencies in the BEC-BCS crossover of a superfluid
Fermi gas.
Superfluidity implies a hydrodynamic behavior which can cause
substantial changes in the excitation spectrum and in general very
low damping rates. However, in the crossover regime the strong
interaction between the particles also results in hydrodynamic
behavior in the normal, non-superfluid phase. Therefore the
interpretation of collective modes in the BEC-BCS crossover in
terms of superfluidity is not straightforward and needs careful
investigation to identify the different regimes.






In this Letter, we report on measurements of fundamental
collective excitation modes in the BEC-BCS crossover for various
coupling strengths in the low-temperature limit. In
Ref.~\cite{li2becinn} we have already presented a first
measurement of the collective excitation of a molecular BEC in the
limit of strong coupling. As described previously
\cite{li2becinn,markus} we work with a spin-mixture of $^6$Li
atoms in the two lowest internal states. For exploring different
interaction regimes we use a broad Feshbach resonance, the
position of which we determined to 837(5)\,G
\cite{exactresonance}. The different interaction regimes can be
characterized by the coupling parameter $1/(k_Fa)$, where $a$
represents the atom-atom scattering length and $k_F$ is the Fermi
wavenumber. Well below the Feshbach resonance ($B<700$\,G) we can
realize the molecular BEC regime with $1/(k_Fa) \gg 1$. On
resonance, we obtain the unitarity-limited regime of a universal
fermionic quantum gas with $1/(k_Fa) = 0$ \cite{heiselberg}. An
interacting Fermi gas of atoms is realized beyond the resonance
where $1/(k_Fa) < 0$.

The starting point of our experiments is a cigar-shaped molecular
BEC 
produced by evaporative cooling in an optical dipole trap in the
same way as described in Ref.~\cite{markus}. Radially the sample
is confined by a 35-mW laser beam (wavelength 1030\,nm) focussed
to a waist of 25\,$\mu$m. The radial vibration frequency is
$\omega_{r} \approx 2\pi\times 750$\,Hz.
The axial vibration frequency is $\omega_z=2\pi \times
(601\,B/$kG$+11)^{1/2}$Hz, where the predominant contribution
stems from magnetic confinement caused by the curvature of the
Feshbach tuning field $B$ and a very small additional contribution
arises from the weak axial optical trapping force.



For exploring collective excitations in the BEC-BCS crossover
regime, we ramp the magnetic field from the evaporation field of
764\,G, where the molecular BEC is formed, to fields between
676\,G and 1250\,G within one second. In previous work
\cite{markus} we have shown that the conversion to an atomic Fermi
gas proceeds in an adiabatic and reversible way, i.e.\ without
increase of entropy. From the condensate fraction in the BEC
limit, for which we measure more than 90\% \cite{markus}, we can
give upper bounds for the temperature in both the BEC limit and
the non-interacting Fermi gas limit of $T<0.46 T_{\rm BEC}$ and
$T<0.03 T_F$ \cite{carr}, respectively. Here $T_{\rm BEC}$ ($T_F$)
denotes the critical temperature (Fermi temperature). With a total
number of atoms $N \approx 4\times10^5$ (free atoms and atoms
bound to molecules) and a geometrically averaged trap frequency at
837G of $\bar{\omega} = (\omega_r^2\omega_z)^{1/3} \approx
2\pi\times 230$\,Hz, we calculate a Fermi energy $E_F = \hbar^2
k_F^2/2m=\hbar \bar{\omega} (3N)^{1/3} =k_B \times 1.2\,\mu$K for
a non-interacting cloud, where
$m$ is the mass of an atom and $k_B$ is Boltzmann's constant.


To excite the {\em axial} compression mode at a given magnetic
field we increase the optical confinement in a 10-ms time interval
by a factor of 1.5. The laser power is varied slow enough for the
radial motion to follow adiabatically, but fast enough to induce
axial oscillations. The relative amplitude of the resulting axial
oscillation is kept small, typically $\sim$10\%. We observe the
oscillation by in-situ imaging of the cloud \cite{markus} after a
variable hold time $t$ at constant trap parameters. To determine
the collective oscillation frequency $\Omega_z$ and the damping
rate $\Gamma_z$ we fit a damped harmonic oscillation $z(t)=z_0 +
A_z \exp(-\Gamma_z t)$sin$(\Omega_z t+\phi_z)$ to the observed
time evolution of the cloud size, where $z_0$, $A_z$, and $\phi_z$
are additional fit parameters.

\begin{figure}
\includegraphics[width=3.3in]{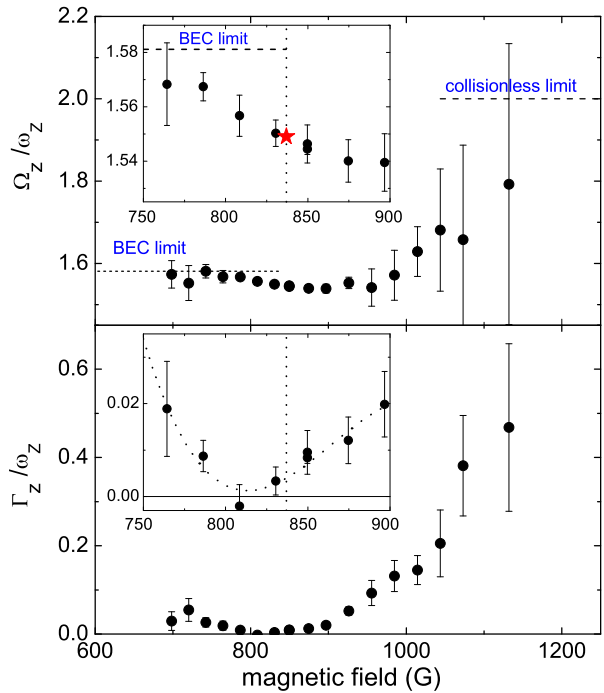}
\caption{Measured frequency $\Omega_z$ and damping rate $\Gamma_z$
of the axial compression mode, normalized to the trap frequency
$\omega_z$. In the upper graph, the dashed lines indicate the BEC
limit of $\Omega_z/\omega_z=\sqrt{5/2}$ and the collisionless
Fermi gas limit with $\Omega_z/\omega_z=2$. The insets show the
data in the resonance region. Here the vertical dotted line
indicates the resonance position at 837(5)\,G. The star marks the
theoretical prediction of $\Omega_z/\omega_z=\sqrt{12/5}$ in the
unitarity limit. In the lower inset the dotted line is a
third-order polynomial fit to the data.} \label{axialosc}
\end{figure}

The measured oscillation frequencies and damping rates are shown
in Fig.~\ref{axialosc}. The data are normalized to the axial trap
frequency $\omega_{z}$, as determined by excitation of the axial
sloshing mode. We point out that the axial confinement is harmonic
because of the dominant magnetic trapping and we can measure
$\omega_z$ with a $10^{-3}$ precision.
In the BEC limit, the measured collective frequencies are in
agreement with the expected $\Omega_z/\omega_{z} =
\sqrt{5/2}=1.581$ \cite{stringari96,ketterleosc}.
With increasing magnetic field we observe a decrease in the
collective excitation frequency until a minimum is reached at
about 900\,G, i.e.\ in the regime of a strongly interacting Fermi
gas where $1/(k_Fa) \approx -0.5$. With further increasing
magnetic field and decreasing interaction strength, we then
observe a gradual increase of the collective frequency toward
$\Omega_z/\omega_z=2$. The latter value is expected for a
collisionless degenerate Fermi gas, where the elastic collision
rate is strongly reduced by Pauli blocking. Because of the large
damping rates in the transition regime between hydrodynamic and
collisionless behavior, the excitation frequencies cannot be
determined with high accuracy. The observed axial damping is
consistent with a gradual transition between these two regimes
\cite{vichi}.

The insets in Fig.~\ref{axialosc} show a zoom-in of the data for
the resonance region between 750 and 900\,G.
The collective frequency that we measure on resonance exhibits the
small 2\% down shift expected for the unitarity limit
($\Omega_z/\omega_{z} = \sqrt{12/5}=1.549$) \cite{stringari04}.
For the damping rate we observe a clear minimum at a magnetic
field of $815(10)\,$G, which is remarkably close to the resonance
position. It is interesting to note that this damping minimum
coincides with the recent observation of a maximum fraction of
condensed fermionic atom pairs in Ref.~\cite{ketterlenew}. For the
minimum damping rate we obtain the very low value of
$\Gamma_z/\omega_{z} \approx 0.0015$, which corresponds to a 1/$e$
damping time of $\sim$5\,s.




\begin{figure}
\includegraphics[width=3in]{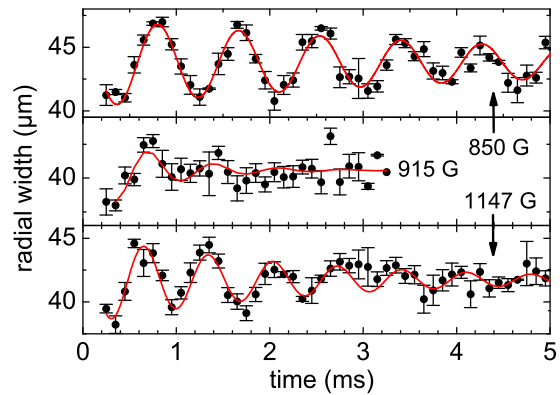}
\caption{Oscillations of the radial compression mode at different
magnetic fields in the strongly interacting Fermi gas regime.
The solid lines show fits by damped harmonic oscillations.}
\label{radialrawcurves}
\end{figure}

To excite the {\em radial} compression mode we reduce the optical
confinement for $50\mu$s, which is short compared with the radial
oscillation period of $1.3\,$ms. In this short interval the cloud
slightly expands radially, and then begins to oscillate when the
trap is switched back to the initial laser power. The relative
oscillation amplitude is $\sim$$10\%$. To detect the radial
oscillation, we turn off the trapping laser after various delay
times $t$ and measure the radial size $r(t)$ after 1.5\,ms of
expansion. The measured radial size $r(t)$ reflects the
oscillating release energy. From the corresponding experimental
data, we extract the excitation frequency $\Omega_r$ and damping
$\Gamma_r$ by fitting the radial cloud size to $r(t)=r_0 + A_r
\exp(-\Gamma_r t)\sin(\Omega_r t+\phi_r)$, where $r_0$, $A_r$ and
$\phi_r$ are additional fit parameters. Typical radial oscillation
curves are shown in Fig.~\ref{radialrawcurves}.

\begin{figure}
\includegraphics[width=3in]{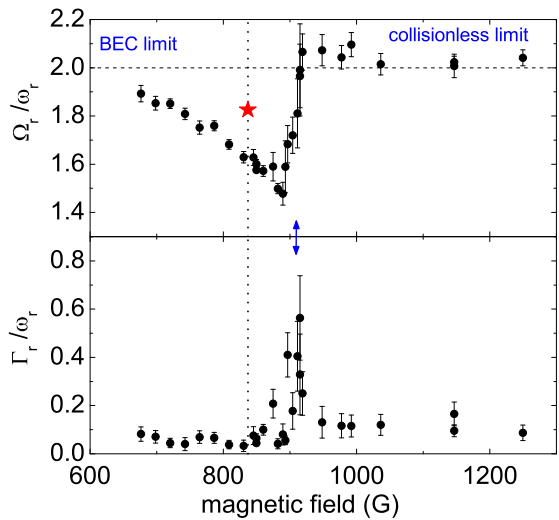}
\caption{Measured frequency $\Omega_r$ and damping rate $\Gamma_r$
of the radial compression mode, normalized to the trap frequency
(sloshing mode frequency) $\omega_r$. In the upper graph, the
dashed line indicates $\Omega_r/\omega_r=2$, which corresponds to
both the BEC limit and the collisionless Fermi gas limit. The
vertical dotted line marks the resonance position at $837(5)$\,G.
The star indicates the theoretical expectation of
$\Omega_r/\omega_r=\sqrt{10/3}$ in the unitarity limit. A striking
change in the excitation frequency occurs at $\sim$910\,G (arrow)
and is accompanied by anomalously strong damping.}
\label{radialosc}
\end{figure}

The magnetic-field dependence of the radial excitation frequency
$\Omega_r$ and the damping rate $\Gamma_r$ is shown in
Fig.~\ref{radialosc}. Here we normalize the data to the trap
frequency $\omega_r$, which we obtain by measuring the radial
sloshing mode at the given magnetic field \cite{radialsloshing}.
This normalization suppresses anharmonicity effects in the
measured compression mode frequency to below 3\%
\cite{anharmonic}. For low magnetic fields, the measured frequency
ratio approaches the BEC limit \cite{stringari96,dalibard}
($\Omega_r/\omega_r = 2$). With increasing magnetic field, i.e.\
increasing interaction strength, we observe a large down-shift of
the frequency. On resonance ($B=837$\,G), we observe
$\Omega_r/\omega_r = 1.62(2)$.
Above resonance, i.e.\ with the gas entering the strongly
interacting Fermi gas regime, the oscillation frequency further
decreases until a maximum shift of almost 30\% ($\Omega_r/\omega_r
= 1.42(5)$) is reached at $B=890\,$G. With further increasing
magnetic field, i.e.\ decreasing interaction strength, an abrupt
change to $\Omega_r/\omega_r \approx 2$ is observed. For
$B>920$\,G our data are consistent with a Fermi gas in the
collisionless regime. The damping of the radial compression mode
is small in the BEC limit and reaches a minimum close to the
unitarity regime. At $B = 910$\,G, where the abrupt change occurs,
we observe very strong damping (see also middle trace in
Fig.~\ref{radialrawcurves}).

We have performed further experiments to check our data on the
radial compression mode for systematic effects. We have repeated
the measurements after recompressing the trap to 9 times higher
trap laser power ($\omega_{r} \approx 2.4\,$kHz). The
corresponding data confirm all our observations in the shallower
trap. In particular, the negative frequency shift and the sudden
change in the collective frequency show up in essentially the same
way. The recompressed trap also allows us to eliminate a small
residual anharmonicity shift from our measurement of the
collective frequency at 837\,G, and we obtain $\Omega_r/\omega_{r}
= 1.67(3)$ for the harmonic trap limit. We have also checked that
the frequency of the compression mode  in the resonance region
does not depend on the way we prepare the ultracold gas. Direct
evaporation at a fixed magnetic field, without starting from a
molecular BEC, leads to the same collective frequency. Preliminary
measurements at higher temperatures, however, show a trend towards
smaller frequency shifts in the radial compression mode and to
smoother changes of the collective frequency.

Our measurements on the radial compression mode show {\em three
surprises}. The corresponding features, which we discuss in the
following, cannot be explained on the basis of available
theoretical models and suggest new physics in the BEC-BCS
crossover regime.

{\em Surprise one:} For a strongly interacting BEC,
Ref.~\cite{pitaevski98} has predicted up-shifts of the collective
frequencies with increasing coupling strength based on beyond
mean-field theory corrections \cite{LHY}. Applying these
predictions to a molecular BEC in the crossover regime, the
collective excitation frequencies should follow
$\delta\Omega_i/\Omega_i=c_i\sqrt{n_{m}a_m^3}$ ($i = z, r$), where
$n_m$ is the peak molecular number density and $a_m=0.6a$
\cite{petrovam} is the molecule-molecule scattering length. For
our highly elongated trap geometry the numerical factors are
$c_r=5c_z=0.727$. In contrast to these expectations, we observe a
strong frequency down-shift in the radial direction. Using the
above formula to fit the first four data points, we obtain a
negative coefficient of $c_r=-1.2(3)$. For the axial oscillation
we obtain $c_z=-0.04(5)$. Note that a substantial down-shift in
radial direction is observed even at the low magnetic field of
676\,G where the molecular gas parameter is relatively small
($n_{m}a_m^3=0.001$). Apparently, the beyond mean field theory of
a BEC is not adequate to describe the transition from a molecular
BEC to a strongly interacting gas in the BEC-BCS crossover.

{\em Surprise two:} The universal character of the strongly
interacting quantum gas on resonance suggests a simple equation of
state for which one expects $\Omega_z/\omega_z=\sqrt{12/5}=1.549$
and $\Omega_r/\omega_r=\sqrt{10/3}=1.826$ for the collective
excitation frequencies \cite{stringari04}. While our measurements
confirm the predicted axial frequency, we obtain a different
frequency in the radial direction of $\Omega_r/\omega_r =
1.67(3)$. 


{\em Surprise three:} The abrupt change of the excitation
frequency and the large damping rate are not expected in a normal
degenerate Fermi gas, where the collective excitation frequency is
expected to vary smoothly from the hydrodynamic regime to the
collisionless one. Furthermore, for the damping rate of the radial
mode in the transition regime a maximum value of
$\Gamma_{r}/\omega_r=0.09$ is calculated in Ref.~\cite{vichi}. Our
measured damping rate of $\Gamma_r/\omega_r \approx 0.5$ is
clearly inconsistent with this prediction for the normal
(non-superfluid) hydrodynamic regime. However, both the sudden
change of the collective frequency and a strong damping are
expected for a transition from the superfluid to the normal phase
\cite{baranov}.

In conclusion, our experiments demonstrate that the collective
modes of a degenerate gas in the BEC-BCS crossover region show a
pronounced dependence on the coupling strength and thus provide
valuable information on the physical behavior of the system. For
the axial compression mode, the frequency shift observed in the
unitarity limit confirms theoretical expectations. However, the
radial compression mode reveals a surprising behavior. In the
strongly interacting BEC regime, the observed frequency shifts
have an opposite sign as compared to expectations from beyond mean
field theory and the frequency shift on resonance is even larger
than expected. The most striking feature is an abrupt change of
the radial collective frequency in the regime of a strongly
attractive Fermi gas where $1/(k_Fa)\approx -0.5$. The transition
is accompanied by very strong damping. The observation supports an
interpretation in terms of a transition from a hydrodynamic to a
collisionless phase.
A superfluid scenario for the hydrodynamic case seems plausible in
view of current theories on resonance superfluidity \cite{holland}
and the very low temperatures provided by the molecular BEC
\cite{carr}. A definite answer, however, to the sensitive question
of superfluidity requires further careful investigations, e.g.\ on
the temperature dependence of the phase transition.

We warmly thank S.\ Stringari for stimulating this work and for
many useful discussions. We also thank W.\ Zwerger and M.\ Baranov
for very useful discussions. We acknowledge support by the
Austrian Science Fund (FWF) within SFB 15 (project part 15) and by
the European Union in the frame of the Cold Molecules TMR Network
under Contract No.\ HPRN-CT-2002-00290. C.C.\ is a Lise-Meitner
research fellow of the FWF.

{\em Note added:} A recent paper by John Thomas's group
\cite{thomasfrequency} reports on measurements of the radial
compression mode in the resonance region, which show much weaker
frequency shifts than we observe. This apparent discrepancy needs
further investigation.


\begin{thebibliography}{10}
\bibitem{radialsloshing}
We excite the radial sloshing mode by pulsed application of a
vertical magnetic gradient. With increasing magnetic field and
thus increasing radial cloud size, the anharmonicity of the
Gaussian-shaped radial potential leads to a decrease of the
sloshing mode frequency. For $\omega_r/2\pi$ we observe a
corresponding decrease from 765\,Hz at 676\,G to 710\,Hz at
magnetic fields above 1.2\,kG; at 837\,G we obtain 740\,Hz. Our
measurements of the sloshing mode frequency agree well with an
expected scaling of the lowest-order anharmonicity correction with
the squared cloud size \cite{stringariprivate}.
\bibitem{anharmonic}
The anharmonicity corrections to frequencies of both the
compression mode ($\Omega_r$) and the sloshing mode ($\omega_r$)
scale in the same way with the increasing cloud size with a
slightly larger prefactor for the compression mode
\cite{stringariprivate}. Our normalization $\Omega_r/\omega_r$
reduces the anharmonicity effect by typically a factor of five.
\end{thebibliography}

\end{document}